\documentclass[amsmath, amssymb, aps, twocolumn]{revtex4-1}
\usepackage{graphicx}
\usepackage{amsmath}
\usepackage{bm}
\usepackage{mathrsfs}
\usepackage{booktabs}

\begin{document}

\title{Surface Shubnikov-de Haas oscillations and non-zero Berry phases of the topological hole conduction in Tl$_{1-x}$Bi$_{1+x}$Se$_2$}

\author{G. Eguchi$^1$}
\email{geguchi@kuee.kyoto-u.ac.jp}
\author{K. Kuroda$^2$} 
\altaffiliation[Current address: ]{Fachbereich Physik und Zentrum f\"ur Materialwissenschaften, Philipps-Universit\"at Marburg, Germany}
\author{K. Shirai$^2$}
\author{A. Kimura$^2$} \author{M. Shiraishi$^1$}
\affiliation{$^1$Department of Electronic Science and Engineering, Graduate School of Engineering, Kyoto University, Kyoto 615-8510, Japan}

\affiliation{$^2$Graduate School of Science, Hiroshima University, Kagamiyama, Higashi-Hiroshima 739-8526, Japan}

\date{\today}

\begin{abstract}
We report the observation of two-dimensional Shubnikov-de Haas (SdH) oscillations in the topological insulator Tl$_{1-x}$Bi$_{1+x}$Se$_2$. Hall effect measurements exhibited electron-hole inversion in samples with bulk insulating properties. The SdH oscillations accompanying the hole conduction yielded a large surface carrier density of $n_{\rm{s}}=5.1 \times10^{12}$/cm$^2$, with the Landau-level fan diagram exhibiting the $\pi$ Berry phase. These results showed the electron-hole reversibility around the in-gap Dirac point and the hole conduction on the surface Dirac cone without involving the bulk metallic conduction.

\begin{description}
\item[PACS numbers]
73.25.+i, 72.20.My, 72.25.-b
\end{description}

\end{abstract}

\maketitle

Electronic and spin transport in the topologically protected surface Dirac cone has been attracting much interest for its potential novel phenomena~\cite{RevModPhys.83.1057,NatNano.9.218}. One of the unique characteristics of the surface metallic state is the lifted spin degeneracy in the presence of time reversal symmetry. In this situation, the surface-Rashba spin-orbit interaction locks the spin direction of the surface electrons. Several spin- and angle-resolved photoemission spectroscopy (spin-ARPES) studies have indeed confirmed this unique spin texture~\cite{PhysRevLett.106.257004}. However, details of the electron and spin transport properties are still unclear. This is because most of the studies on topological insulators reported to date also involve bulk metallic conduction~\cite{doi:10.7566/JPSJ.82.102001}, making it difficult to separate the surface metallic conduction.

TlBiSe$_2$ is known to exhibit an in-gap Dirac point, where both the lower and upper parts of the surface Dirac cone are confined in a bulk energy gap, which is missing in the well-studied Bi$_2$Se$_3$~\cite{RevModPhys.83.1057,PhysRevLett.105.036404,PhysRevLett.105.136802,PhysRevLett.105.146801}. Recently, Fermi-level tuning as well as bulk insulating and surface metallic conduction have been achieved in Tl$_{1-x}$Bi$_{1+x}$Se$_2$~\cite{arXiv.1308.5521}. These features are a great advantage for a future ambipolar gate control and a spin-transport with the Dirac cone, thus investigations of the transport properties with this system are of notable significance.

In this article, we report the observation of two-dimensional Shubnikov-de Haas (SdH) oscillations in Tl$_{1-x}$Bi$_{1+x}$Se$_2$, finding the hole conduction with bulk insulating behavior. The Landau-level fan diagram obtained from the oscillations exhibited a phase shift of $\pi$ because of the finite Berry phase. Inversion of the electron-hole conduction was also observed in samples of bulk insulator. Thus far, a surface metallic conduction with electrons, as well as bulk insulating behavior has been achieved in Bi$_2$Te$_3$, Bi$_2$Te$_2$Se and Bi$_{2-x}$Sb$_{x}$Te$_{3-y}$Se$_{y}$~\cite{Qu29072010,PhysRevB.82.241306,PhysRevB.86.045314,PhysRevLett.107.016801}. The surface hole conduction was also reported in the Bi$_{2-x}$Sb$_{x}$Te$_{3-y}$Se$_{y}$, however it dissipates by time and an additional metallic conduction originating from the bulk is also involved~\cite{PhysRevLett.107.016801}. The success of the surface conduction of holes and the electron-hole inversion in the bulk-insulator Tl$_{1-x}$Bi$_{1+x}$Se$_2$ in the no-dissipative condition offers a new opportunity for manipulating the surface transport.

\begin{figure}[!h]
\begin{center}
\includegraphics[width=\columnwidth,clip]{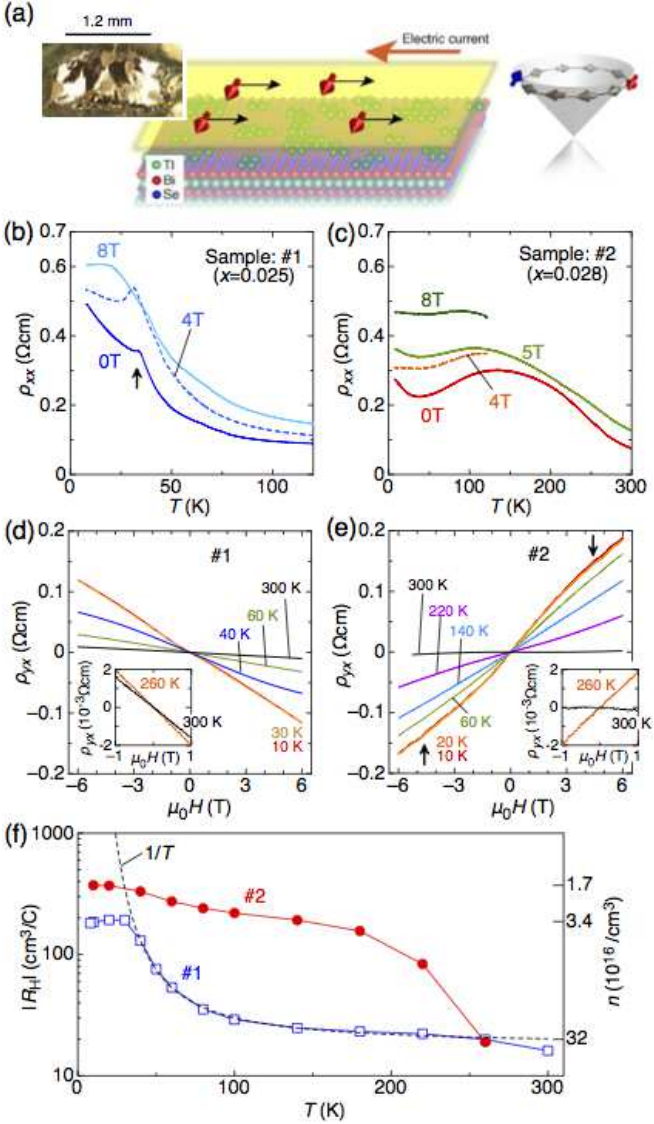}
\caption{(Color online) (a)~Schematics of the surface electron conduction on the cleaved surface of Tl$_{1-x}$Bi$_{1+x}$Se$_2$. The thallium islands are formed on the surface~\cite{PhysRevB.88.245308}. The surface Dirac cone with the spin helical texture, and a photo of sample \#2 are also presented. (b) and (c)~Temperature dependence of the resistivity ($\rho_{xx}$) in each sample, down to 8~K under several different magnetic fields. The $\rho_{xx}$ at 0~T for \#1 is the same data presented in Fig.~3(c) of Ref.~\cite{arXiv.1308.5521}. (d) and (e)~Magnetic field dependence of the Hall resistivity ($\rho_{yx}$) in each sample. The insets show the $\rho_{yx}$ values near room temperature. The arrows in (e) indicate where the Shubnikov-de Haas oscillation occurred. (f)~Temperature dependence of the Hall coefficient ($|R_{\rm{H}}|$) in each sample. The corresponding bulk carrier density ($n$) is also presented. The dashed curve is the Arrhenius plot, ${\rm{log}}|R_{\rm{H}}| \propto 1/T$, which represents the bulk insulating behavior.}
\label{fig1}
\end{center}
\end{figure}

The Tl$_{1-x}$Bi$_{1+x}$Se$_2$ single crystals used in this study were synthesized by the Bridgman technique~\cite{arXiv.1308.5521}. The two samples reported in this Letter are identified as \#1 ($x=0.025$) and \#2 ($x=0.028$) with the $x$ values determined using electron probe micro-analysis. The \#1 was from the same batch of samples as that reported in Ref.~\cite{arXiv.1308.5521}. Electric transport measurements were performed using the conventional 6-probe technique down to 8~K with a homebuilt probe assembled for a commercial apparatus (Quantum Design PPMS). The electric contacts of both samples were made with a room-temperature-cured silver paste that was put onto the cleaved surface (see Fig.~\ref{fig1}(a)). The whole process for the contacts was carried out in the air. The sample thicknesses were 0.46~mm (\#1) and 0.21~mm (\#2). 

The temperature dependence of the resistivity ($\rho_{xx}$) under different magnetic fields is presented in Fig.~\ref{fig1}(b) for \#1 and Fig.~\ref{fig1}(c) for \#2. The $\rho_{xx}$ values for \#1 exhibited the negative temperature coefficient, indicating the bulk insulating behavior. A weak saturation below 40~K was also observed as indicated by an arrow. The $\rho_{xx}$ values for \#2 exhibited the negative temperature coefficient above 150~K, and significantly, the coefficient turned into positive below the temperature indicating the metallic behavior. These characteristic temperature dependences of the $\rho_{xx}$ were consistent with other bulk-insulating topological insulators~\cite{PhysRevLett.107.016801,PhysRevB.85.155301} which suggests the presence of the surface metallic conduction at low-temperatures. The positive temperature dependence became unclear under a magnetic field, and the $\rho_{xx}$ values at low-temperatures exhibited a weak temperature dependence. The weak temperature dependence of the $\rho_{xx}$ coincides with the surface metallic behavior~\cite{PhysRevB.82.241306,PhysRevB.86.045314,PhysRevLett.107.016801,PhysRevB.85.155301}. As shown in Fig.~\ref{fig1}(b), the weak saturation for \#1 in magnetic fields behaved in a similar manner to the metallic behavior of \#2. The result strongly suggests that the saturation originated from the surface metallic conduction.

\begin{figure}[!h]
\begin{center}
\includegraphics[width=\columnwidth,clip]{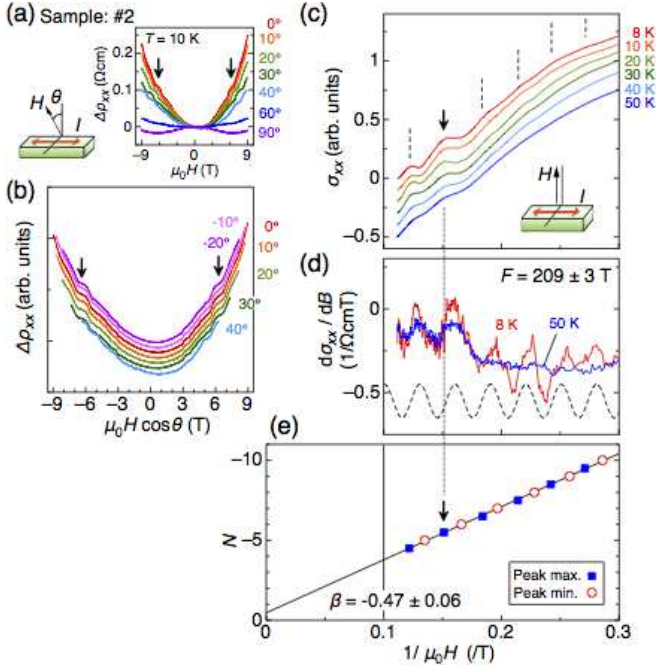}
\caption{(Color online) (a)~Magnetic field dependence of the $\varDelta\rho_{xx}=\rho_{xx}(B)-\rho_{xx}(0)$ for the sample \#2 at 10~K for several different field angles ($\theta$). A schematic of the measurement setup is also presented. The weak anti-localization expected in the system was not observed within our experimental resolution~\cite{doi:10.7566/JPSJ.82.102001}. (b)~The $\mu_0 H \perp (001)=\mu_0 H{\rm{cos}}\theta$ dependence of the $\varDelta\rho_{xx}$. (c), (d) and (e)~Inverse magnetic field ($1/\mu_0 H$) dependence of (c)~the $\sigma_{xx}=1/\rho_{xx}$ with a constant offset at several different temperatures, (d)~${\rm{d}}\sigma_{xx}/{\rm{d}}B$ at 8~K and 50~K, and (e)~the Landau-level fan diagram determined from the $\sigma_{xx}$ in 8~K. The vertical dashed lines in (c) represent the SdH peaks. A schematic of the measurement setup is also presented in (c). The dashed curve in (d) is the fitting for the 8~K data with a sine curve, giving a frequency $F=209 \pm3$~T. See the text for details. The solid line in (e) is the fit of $N=F/(2\pi B_N)+\beta$, giving $\beta=-0.47 \pm 0.06$. The arrows in each figure indicate the same SdH field, corresponding to $N=-5.5$.}
\label{fig2}
\end{center}
\end{figure}

The magnetic field dependence of the Hall resistivity ($\rho_{yx}$) at several temperatures is presented in Fig.~\ref{fig1}(d) for \#1 and Fig.~\ref{fig1}(e) for \#2. The insets show the $\rho_{yx}$ near room temperature. As shown in Fig.~\ref{fig2}(d), the $\rho_{yx}$ values for \#1 exhibited a negative slope for each temperature, indicating that electron conduction was dominant. In contrast, the $\rho_{yx}$ values for \#2 exhibited a positive slope below 260~K, indicating that hole conduction was dominant. According to the ARPES results~\cite{arXiv.1308.5521}, the surface conduction of electrons was expected for both samples because the Fermi levels ($E_{\rm{F}}$) were higher than the energy of the Dirac point ($E_{\rm{DP}}$): $E_{\rm{F}}>E_{\rm{DP}}$. This surprising result of the positive slope suggests that $E_{\rm{F}}$ for \#2 has shifted to an energy of $E_{\rm{F}}<E_{\rm{DP}}$, and the surface conduction of holes. It is noteworthy that the surface conduction of electrons inverted from holes by exposure to air is reported in other topological insulators~\cite{PhysRevLett.107.016801}. The fact hints the surface conduction of holes inverted from electrons for \#2 by exposure to air. As shown in the inset of Fig.~\ref{fig1}(e), the positive slope of the $\rho_{yx}$ diminished at 300~K. This suggests presence of thermally-activated additional carriers presumably due to the bulk conduction. The arrows in Fig.~\ref{fig1}(e) notify the oscillatory behavior that was caused by the SdH effect. Observation of the SdH oscillations confirmed the presence of the metallic state, since the oscillatory behavior originates from the Landau quantization of the finite density of states at $E_{\rm{F}}$. This behavior will be discussed later. 

Now, we examine the temperature dependence of the Hall coefficient ($|R_{\rm{H}}|$) calculated from the relation: $R_{\rm{H}}=\rho_{yx}/(\mu_0 H)$. Since the low-field $\rho_{yx}$ in each temperature was well explained by the linear field dependence over the whole temperature range, the data at $\pm1$~T were used for further analyses. The $|R_{\rm{H}}|$ values at 10~K were 190~cm$^3$/C for \#1 and 370~cm$^3$/C for \#2. The corresponding carrier densities ($n$) were calculated from these values with the relation: $R_{\rm{H}}=-1/(nq)$, yielding 3.4$\times10^{16}$~/cm$^3$ for \#1 and 1.7$\times10^{16}$~/cm$^3$ for \#2. Here $q$ is the charge and is expressed as $q=+e$ for electrons or $q=-e$ for holes where $e$ is the elementary charge. As shown in Fig.~\ref{fig1}(f), the $|R_{\rm{H}}|$ for \#1 exhibited a small temperature dependence below 30~K and that for \#2 below 200~K. Generally, the carrier density of a metal has a small temperature dependence, whereas that of a semiconductor follows the Arrhenius law: ${\rm{log}}(n) \propto -1/T$. In fact, above 30~K the $|R_{\rm{H}}|$ for \#1 agreed well with the Arrhenius law, as indicated with the dashed curve in Fig.~\ref{fig1}(f). Thus, the observed small temperature dependence below 30~K for \#1 and 200~K for \#2 were most consistent with the metallic behavior.

The Hall mobilities ($\mu_{\rm{H}}$) deduced from the relation: $\mu_{\rm{H}}=|R_{\rm{H}}|/\rho_{xx} (0)$ were 390~(10~K) to 200~(300~K)~cm$^2$/Vs for \#1 and 1400~(10~K)~ to ~140~(260~K)~cm$^2$/Vs for \#2, where $\rho_{xx}(0)$ is the resistivity at 0~T. The large enhancement of $\mu_{\rm{H}}$ observed in \#2 suggested the larger contribution of the surface conduction.

Next, we focus on the SdH effect of \#2 as mentioned with Fig.~\ref{fig1}(e) and discuss the surface metallic behavior. The magnetic field dependence of the $\varDelta\rho_{xx}=\rho_{xx}(B)-\rho_{xx}(0)$ with $B=\mu_0 H$ at 10~K for several field angles ($\theta$) is presented in Fig.~\ref{fig2}(a). As indicated with the arrows, a clear shift in the oscillations by $\theta$ was observed. The oscillations were well defined with the field strength perpendicular to the cleaved surface: $\mu_0 H \perp (001)=\mu_0 H{\rm{cos}}\theta$, as presented in Fig.~\ref{fig2}(b). The Fig.~\ref{fig2}(b) shows $\varDelta\rho_{xx}$ as a function of $\mu_0 H {\rm{cos}}\theta$. The result indicates that the oscillations have the two-dimensional nature and therefore confirms the presence of the surface metallic state.

The inverse magnetic field ($1/\mu_0 H$) dependence of the conductivity ($\sigma_{xx}=1/\rho_{xx}$) with a constant offset is presented in Fig.~\ref{fig2}(c) for several different temperatures. The arrows in Figs.~\ref{fig2}(b) and (c) indicate the same SdH field. The derivation of $\sigma_{xx}$ with $B$: ${\rm{d}}\sigma_{xx}/{\rm{d}}B$ at 8~K and 50~K is presented in Fig.~\ref{fig2}(d). The 8~K data were fitted with ${\rm{d}}\sigma_{xx}/{\rm{d}}B \propto -{\rm{sin}}(F/B+\varphi)$, where $F$ is the frequency and $\varphi$ is the phase resulting in $F=209 \pm 3$~T and $\varphi=(0.09 \pm0.04) \pi$. Obviously, the oscillations were well defined with a single frequency which arose from the the single Fermi surface lying at the top surface. The Landau-level fan diagram indexed by the relation: $N=F/(2\pi B_N)+\beta$ (solid line) is presented in Fig.~\ref{fig2}(e). By considering the conduction of holes, the $\beta$ value with negative peak indices is to be $-1/2$ owing to the $\pi$ Berry phase. As indicated in the figure, fitting the peak indices resulted in $\beta=-0.47 \pm 0.06$. The result of $\beta$ value, close to $-1/2$, indicates the oscillations originate from the surface helical state. Note that $\beta$ has a relation $\varphi=2\pi(\beta-1/2)$, and the $\varphi$ value obtained was also consistent with the $\pi$ Berry phase.

The cross section of the Fermi surface ($A$) was determined using the Onsager relation: $F=\hbar A/2\pi e$. Since the deformation of the Fermi surface in TlBiSe$_2$ is known to be small~\cite{PhysRevLett.106.216803}, the circular assumption with the relationships $A=\pi k_{\rm{F}}^2$ and $A=(2\pi)^2n_{\rm{s}}$ can be applied. Here $\hbar$ is the reduced Planck constant, $k_{\rm{F}}$ is the Fermi wave number, and $n_{\rm{s}}$ is the surface carrier density. The $A$ yielded $k_{\rm{F}}=8.0 \times 10^{6}$~/cm and $n_{\rm{s}}=5.1 \times 10^{12}$~/cm$^2$. The $k_{\rm{F}}$ value is compared with the ARPES results~(see the ARPES spectrum presented in Fig.~2(e) in Ref.~\cite{PhysRevLett.105.146801}). By considering the relation $E_{\rm{F}}<E_{\rm{DP}}$, the $k_{\rm{F}}$ value obtained corresponded to $E_{\rm{F}}-E_{\rm{DP}}=-0.20$~eV. The Fermi velocity ($v_{\rm{F}}$) was also obtained as $v_{\rm{F}}=4.1\times 10^7$~cm/s, which was close to the  velocity at $E_{\rm{DP}}$: $v_{\rm{DP}}=3.9\times 10^7$~cm/s~\cite{PhysRevLett.105.146801}. The fact confirms the Dirac surface state origin of the measured SdH oscillation and indicates the small distortion from the linear dispersion relation: $E(k)=E_{\rm{DP}}+v_{\rm{DP}}\hbar k$ at the $k_{\rm{F}}$.

\begin{figure}[!h]
\begin{center}
\includegraphics[width=\columnwidth,clip]{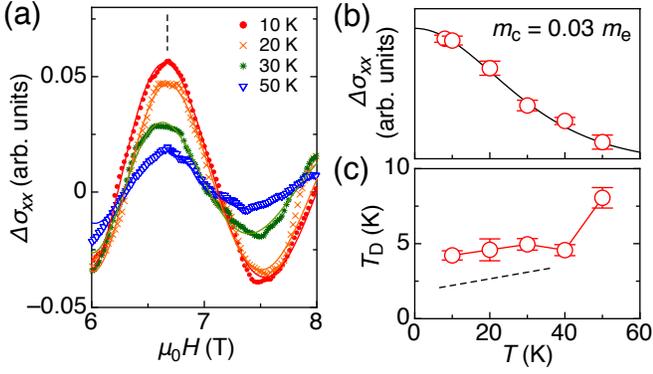}
\caption{(Color online)~(a)~Oscillatory component of the $\sigma_{xx}$, deduced from the data in Fig.~\ref{fig2}(c). The Solid lines show the fit with the standard Lifshitz-Kosevich (LK) theory~\cite{PhysRevB.86.045314}. The peak corresponding $N=-5.5$ is indicated by a dashed line. (b) Temperature dependence of the oscillation amplitude of the peak. The solid line is the fit with the LK theory, resulting in a cyclotron mass $m_{\rm{c}}=(0.03 \pm 0.01)m_{\rm{e}}$. (c) Temperature dependence of the Dingle temperature ($T_{\rm{D}}$). Each  $T_{\rm{D}}$ value was deduced from the data presented in Fig.~\ref{fig3}(a). The $T_{\rm{D}}$ at 10~K was $4.2 \pm 0.3$~K. The dashed line indicates a linear temperature dependence of $T_{\rm{D}}$ observed below 40~K.}
\label{fig3}
\end{center}
\end{figure}

For further analysis of the surface hole-Dirac fermions, the oscillatory component of $\sigma_{xx}$ was deduced from the data in Fig.~\ref{fig2}(c) by subtracting the 2nd polynomial fit for $\sigma_{xx}$. The resultant oscillatory component of $\sigma_{xx}$ at several different temperatures is presented in Fig.~\ref{fig3}(a). The fits with the standard Lifshitz-Kosevich (LK) theory are also presented with solid lines~\cite{PhysRevB.86.045314}. The temperature dependence of the oscillation amplitude, deduced from the peak at $N=-5.5$, is presented in Fig.~\ref{fig3}(b). The solid line is the fit using the LK theory under a constant magnetic field: $\varDelta  \sigma_{xx} \propto \lambda/{\rm{sinh}}\lambda$ with $\lambda=2\pi^2m_{\rm{c}}k_{\rm{B}}T/(\hbar eB)$, where $m_{\rm{c}}$ is the cyclotron mass and $k_{\rm{B}}$ is the Boltzmann constant. The fitting results for $m_{\rm{c}}=(0.03 \pm 0.01)m_{\rm{e}}$, where $m_{\rm{e}}$ is the electron rest mass. The Dingle temperature ($T_{\rm{D}}$) at each temperature was also deduced from the fitting, yielding $T_{\rm{D}}=4.2 \pm0.3$~K at 10~K. A positive temperature coefficient of $T_{\rm{D}}$ was observed, as indicated in the figure. The $T_{\rm{D}}$ determined the lower bound of the lifetime $\tau=\hbar/(2\pi k_{\rm{B}} T_{\rm{D}})$~\cite{Ferry_Goodnick_Bird}, and the linear temperature dependence of $T_{\rm{D}}$ implies $1/\tau \propto T$ which is consistent with a metallic behavior. The $T_{\rm{D}}$ value at 10~K yielded $\tau = 2.9 \times10^{-13}$~s, surface mean free path $l=v_{\rm{F}}\tau=120$~nm, and mobility $\mu_{\rm{s}}=el/(\hbar k_{\rm{F}})=2200$~cm$^2$/Vs of the surface holes. The $\mu_{\rm{s}}$ in addition to the $n_{\rm{s}}$ yielded the sheet conductance $G_{\rm{s}}=n_{\rm{s}}e \mu_{\rm{s}} = 1.8 \times 10^{-3}$~/$\Omega$, and the $G_{\rm{s}}$ determined the lower bound of the surface contribution to the total conductivity as $G_{\rm{s}}/\sigma_{xx}d=2.3$\%. Note that $d=0.21$~mm is the sample thickness. Taking the large sample thickness and the high temperature into account, the obtained surface contribution is considered to be large compared with other topological insulators~\cite{Qu29072010,PhysRevB.82.241306,PhysRevLett.107.016801}.

\begin{figure}[!h]
\begin{center}
\includegraphics[width=\columnwidth]{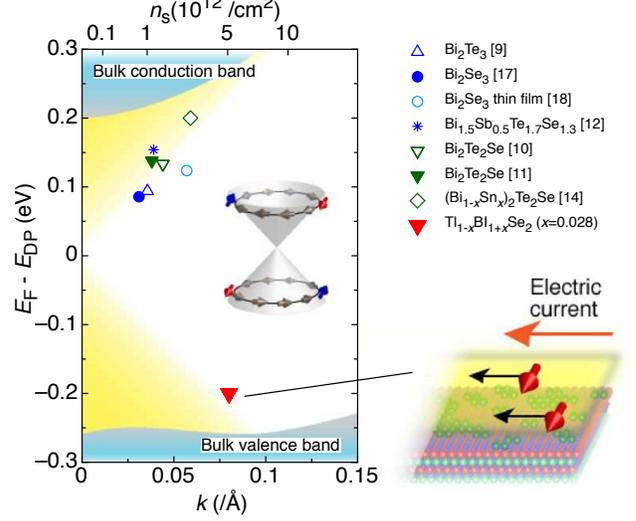}
\caption{(Color online)~The Fermi level ($E_{\rm{F}}-E_{\rm{DP}}$) determined from the Shubnikov de-Haas oscillations of known three-dimensional topological insulators~\cite{Qu29072010,NatPhys.6.960,PhysRevB.82.241306,PhysRevLett.107.016801,PhysRevB.86.045314,PhysRevLett.109.066803,PhysRevB.85.155301}, where $E_{\rm{F}}$ is the Fermi level, and $E_{\rm{DP}}$ is the energy of the Dirac point. The surface carrier density ($n_{\rm{s}}$) is also presented. Schematics of the surface Dirac cone and the topological hole conduction on a TlBiSe$_2$ crystal is also presented.}
\label{fig4}
\end{center}
\end{figure}

Finally, the $E_{\rm{F}}-E_{\rm{DP}}$ values determined from the SdH oscillations, of known three-dimensional topological insulators are summarized in Fig.~\ref{fig4}~\cite{NatPhys.6.960,PhysRevB.82.241306,PhysRevLett.107.016801,PhysRevB.86.045314,PhysRevLett.109.066803,PhysRevB.85.155301}. Thus far, Tl$_{1-x}$Bi$_{1+x}$Se$_2$ appears to exhibit the largest $n_{\rm{s}}$ reported. This implies the comparably large $G_{\rm{s}}/\sigma_{xx}d$ value obtained is ascribable to the large $n_{\rm{s}}$. As shown in the figure, this is also the only case where the $E_{\rm{F}}$ stabilized in $E_{\rm{F}}<E_{\rm{DP}}$. The fact indicates that the spins are carried by the holes (see the schematic in Fig.~\ref{fig4}). We also refer to the sign inversion for $R_{\rm{H}}$ in samples accompanied by the bulk insulating, surface metallic behavior. These results on electric transport confirmed the electron-hole reversibility of the in-gap Dirac point, and are consistent with the ARPES results~\cite{arXiv.1308.5521}.

In summary, we investigated the magnetic field dependence of the electric transport in Tl$_{1-x}$Bi$_{1+x}$Se$_2$. The surface SdH oscillations were observed in a sample with bulk insulating, surface metallic conduction of holes. By comparing the obtained $k_{\rm{F}}$ value of $8.0 \times 10^{6}$~/cm with the ARPES spectrum, the Fermi level of the sample was determined to be $E_{\rm{F}}-E_{\rm{DP}}=-0.20$~eV. The temperature dependence observed for the Hall coefficient coincided with the surface conduction. It should be emphasized that TlBiSe$_2$ exhibits a simple Dirac cone with an in-gap Dirac point. Therefore, this is suitable for future transport studies such as gate tuning and the spin-injection.

We thank Yoshifumi Ueda, Michael Delmo and Yuuichiro Ando for their fruitful discussions and useful advice. This work was supported by a Grant-in-Aid for Scientific Research from the Ministry of Education, Culture, Sports, Science and Technology of Japan, and Innovative Areas "Nano Spin Conversion Science". G.E. and K. K. were supported by the Japan Society for the Promotion of Science (JSPS).

\hspace{1pt}

\end{document}